\documentclass[acmsmall,authorversion]{acmart}

\usepackage{algorithm}
\usepackage{algpseudocode}
\usepackage[export]{adjustbox}
\usepackage{tikz}
\RequirePackage{luatex85,shellesc}
\usetikzlibrary{arrows.meta,positioning,calc,shapes.misc} 
\usepackage{pgfplots}
\usepackage{pgfplotstable}
\pgfplotsset{compat=1.18}
\AtBeginDocument{%
  }

\setcitestyle{square}

\setcopyright{acmlicensed}
\acmJournal{PACMCGIT}
\acmYear{2024} \acmVolume{7} \acmNumber{1} \acmArticle{} \acmMonth{5}\acmDOI{10.1145/3651298}

\begin{document}

\title{Transforming a Non-Differentiable Rasterizer into a Differentiable One with Stochastic Gradient Estimation}

\author{Thomas Deliot}
\email{thomas.deliot@intel.com}

\author{Eric Heitz}
\email{eric.heitz@intel.com}

\author{Laurent Belcour}
\affiliation{%
\institution{Intel Corporation}
\city{Grenoble}
\country{France}}
\email{laurent.belcour@intel.com}

\renewcommand{\shortauthors}{Deliot et al.}

\begin{abstract}
We show how to transform a non-differentiable rasterizer into a differentiable one with minimal engineering efforts and no external dependencies (no Pytorch/Tensorflow).
We rely on \emph{Stochastic Gradient Estimation}, a technique that consists of rasterizing after randomly perturbing the scene's parameters such that their gradient can be stochastically estimated and descended.
This method is simple and robust but does not scale in dimensionality (number of scene parameters). 
Our insight is that the number of parameters contributing to a given rasterized pixel is bounded.
Estimating and averaging gradients on a per-pixel basis hence bounds the dimensionality of the underlying optimization problem and makes the method scalable. 
Furthermore, it is simple to track per-pixel contributing parameters by rasterizing ID- and UV-buffers, which are trivial additions to a rasterization engine if not already available. 
With these minor modifications, we obtain an in-engine optimizer for 3D assets with millions of geometry and texture parameters. 
\end{abstract}



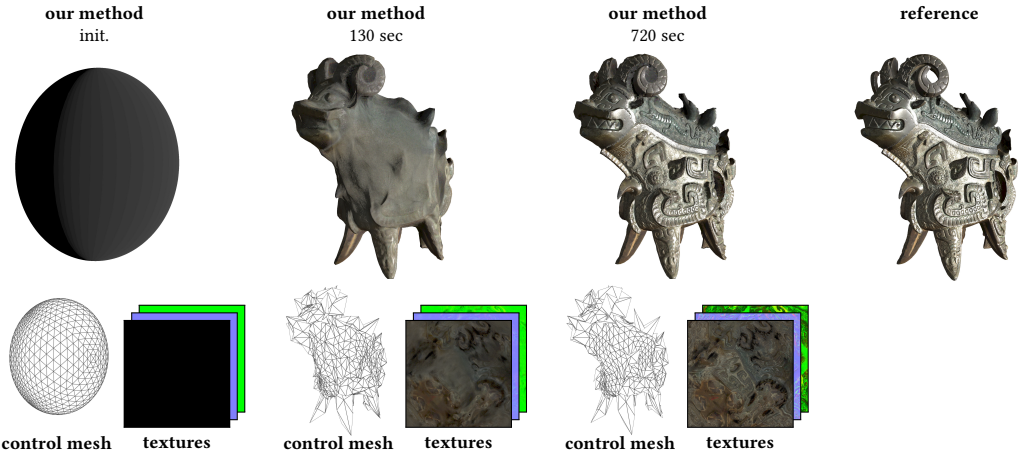
\begin{teaserfigure}
\vspace{-3mm}
\begin{tabular}{@{\hspace{-5mm}}c@{}}
\begin{tikzpicture}
\begin{scope}[shift={(0,0)}]
\draw (0, 0) node {\includegraphics[width=0.25\linewidth]{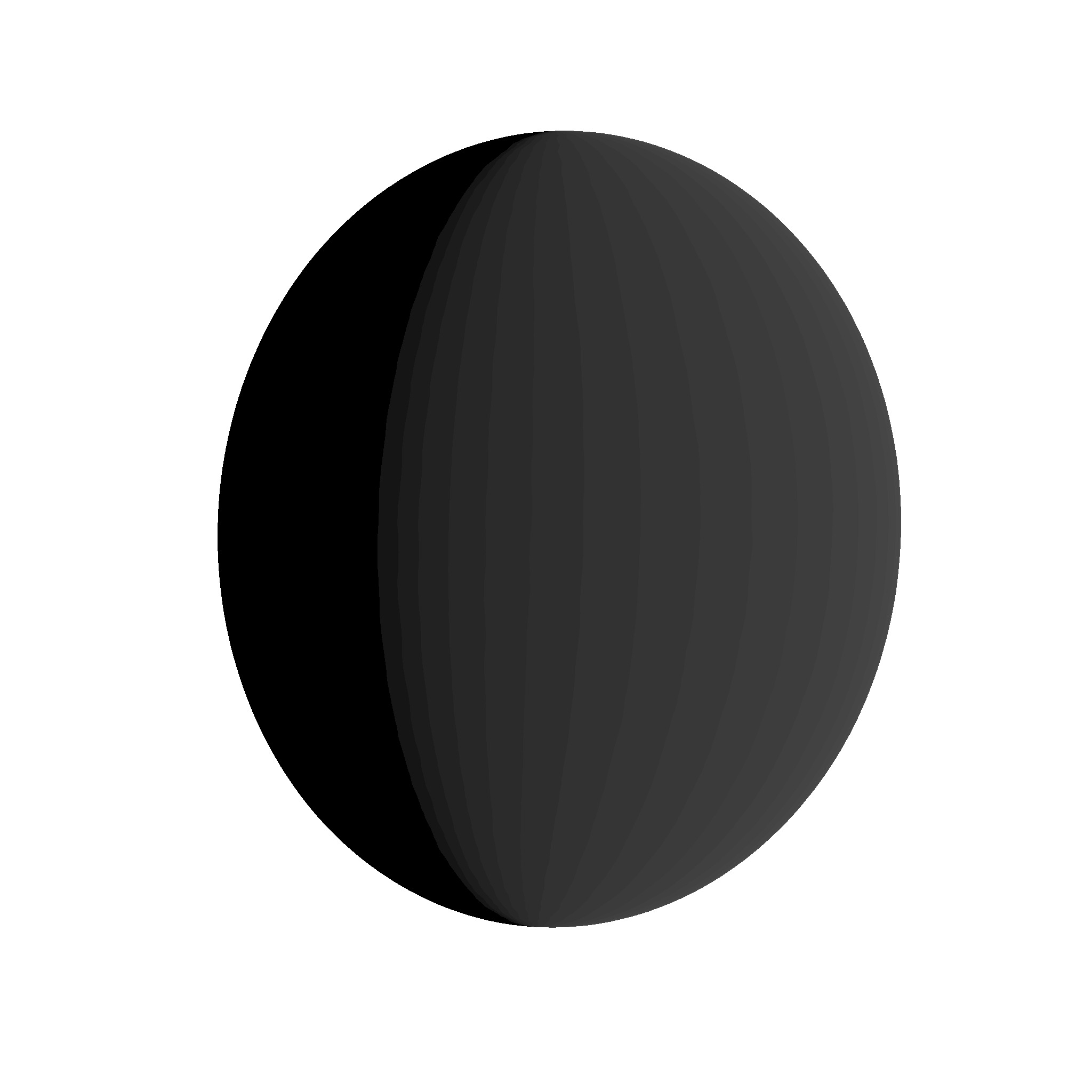} };
\draw (-0.5,-2.5) node {\includegraphics[width=0.15\linewidth]{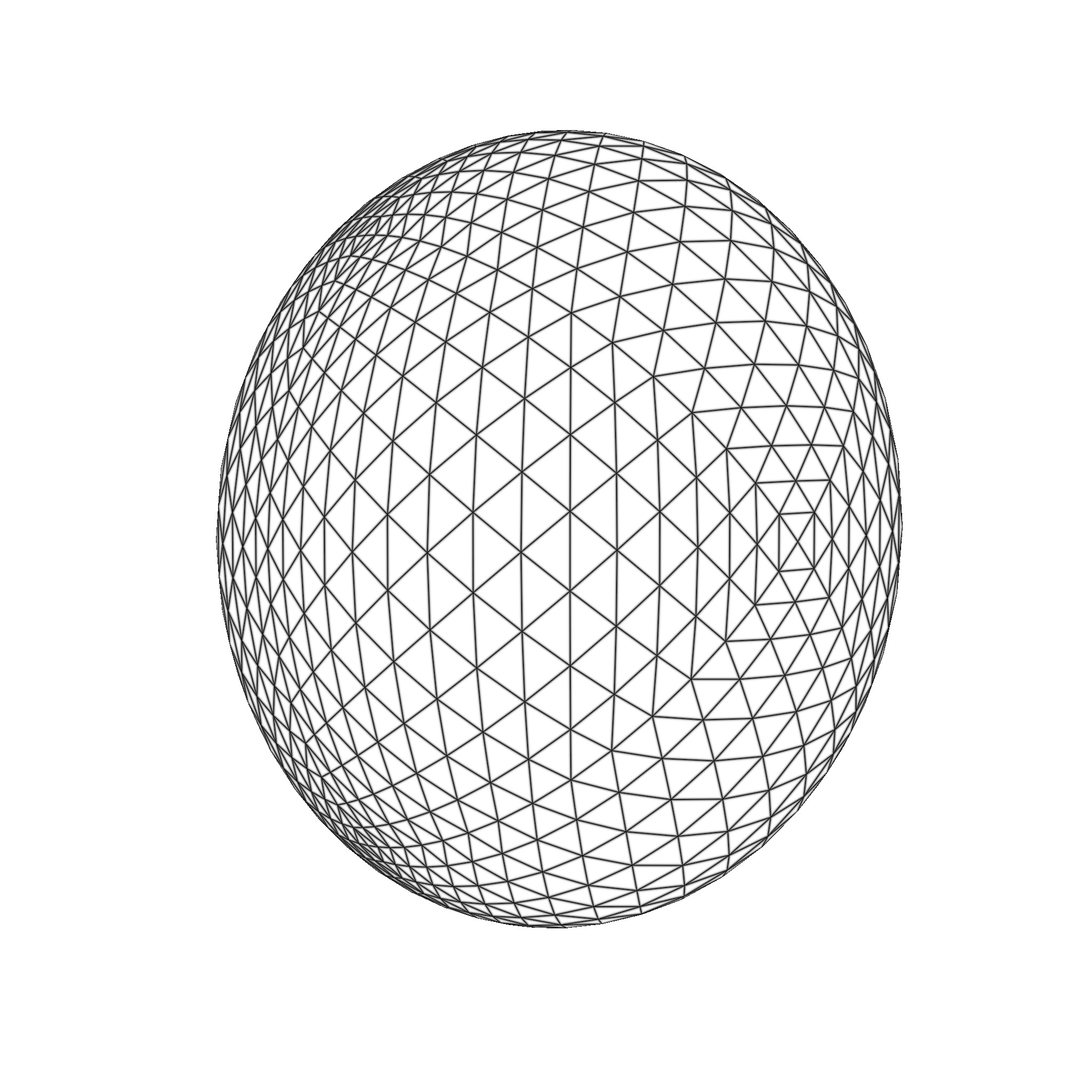} };
\draw (1.1 + 2*0.1, -2.5 - 0*0.1) node {\includegraphics[width=0.1\linewidth, frame]{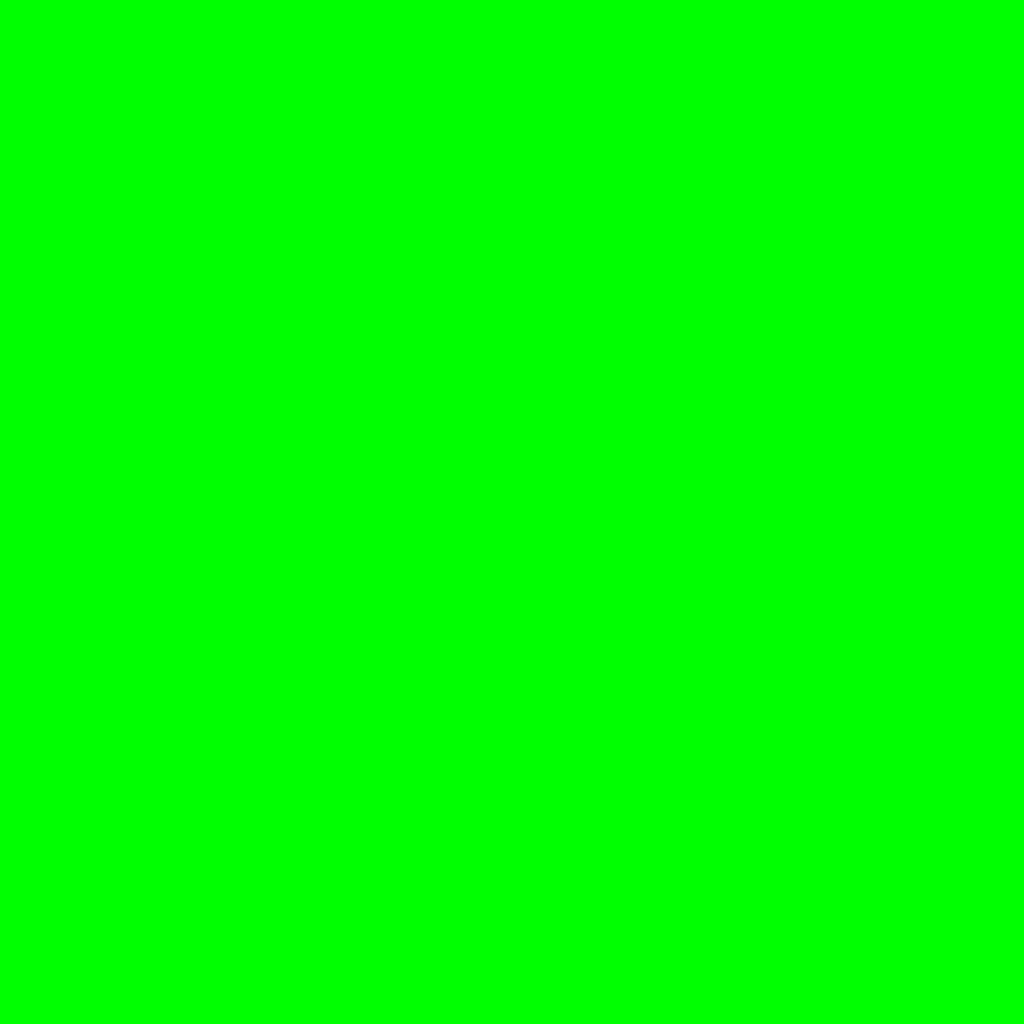} };
\draw (1.1 + 1*0.1, -2.5 - 1*0.1) node {\includegraphics[width=0.1\linewidth, frame]{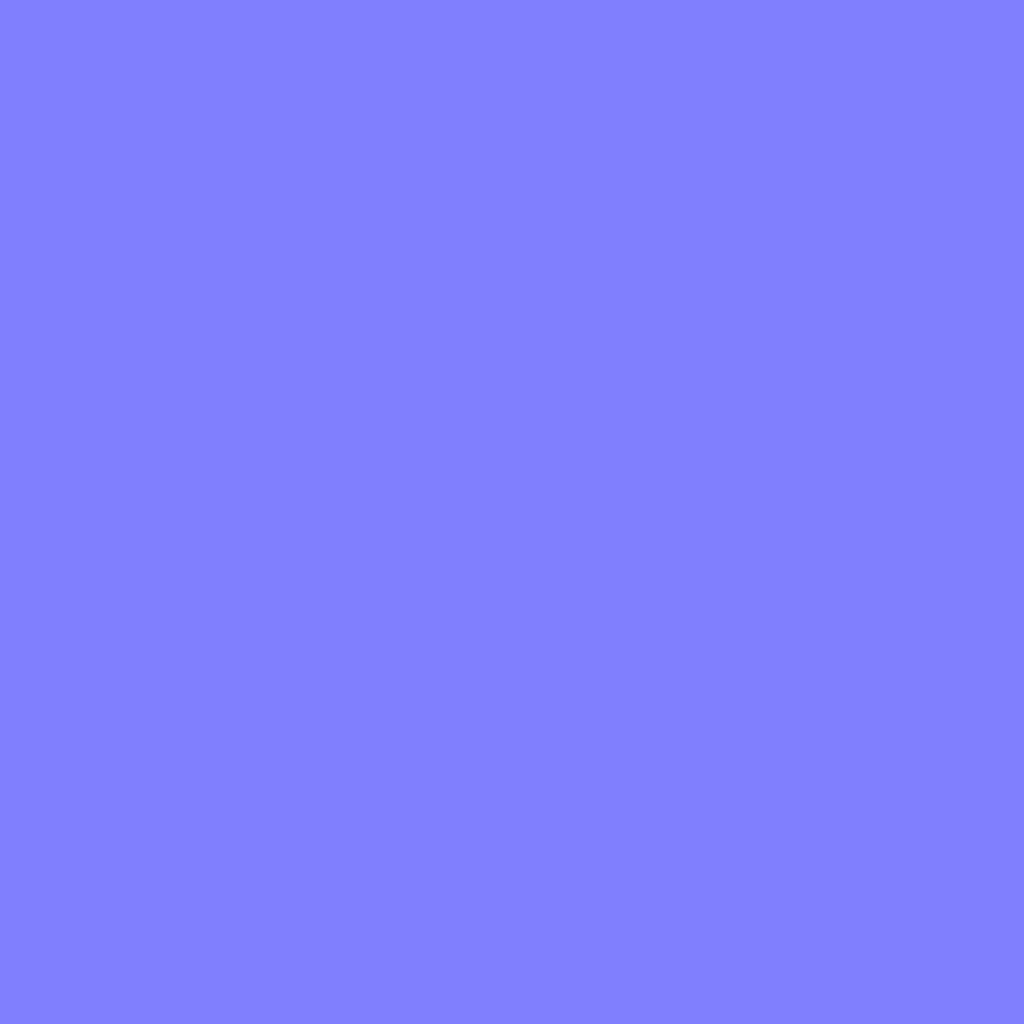} };
\draw (1.1 + 0*0.1, -2.5 - 2*0.1) node {\includegraphics[width=0.1\linewidth, frame]{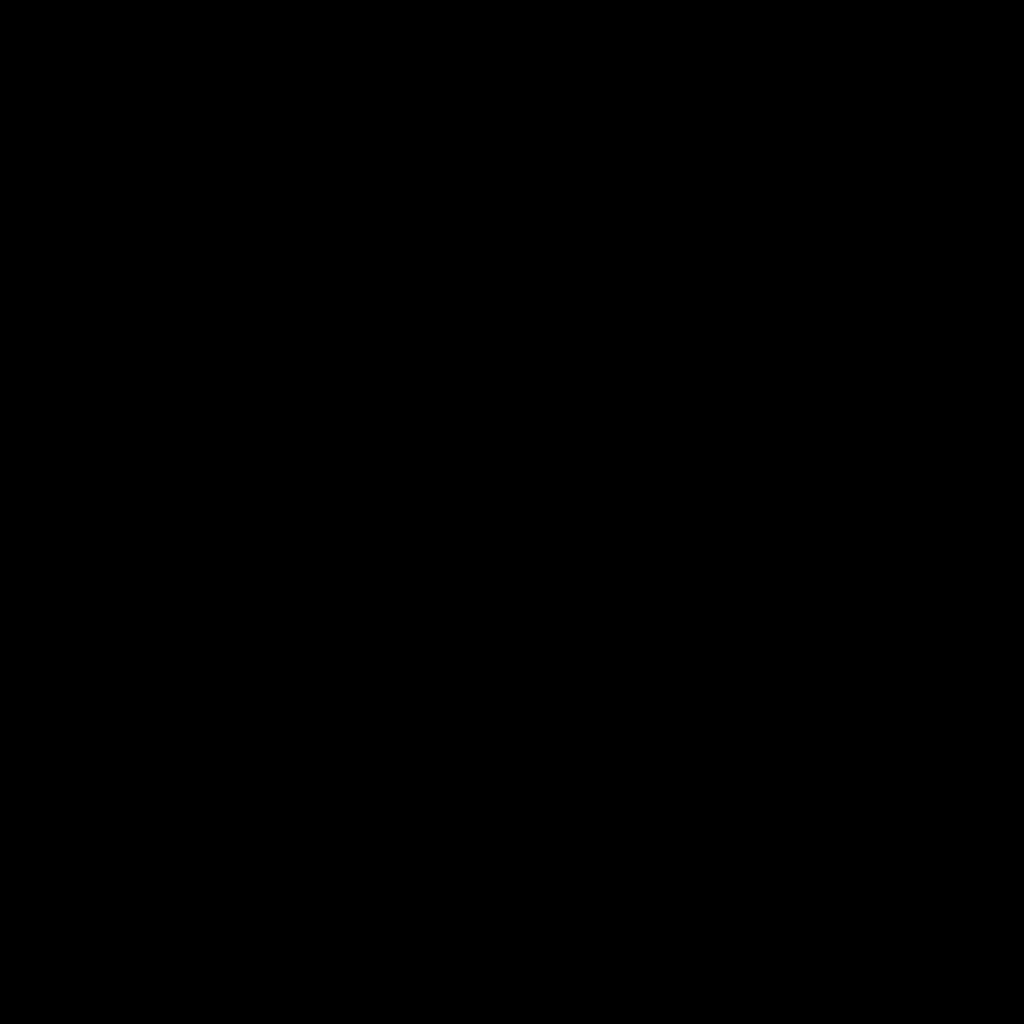} };
\draw (-0.5, -3.6) node {\scriptsize \textbf{control mesh}};
\draw (+1.0 + 0.1, -3.6 - 0.01) node {\scriptsize \textbf{textures}};
\draw (0, 1.75) node {\scriptsize {init.}};
\draw (0, 2.05) node {\scriptsize \textbf{our method}};
\end{scope}
\begin{scope}[shift={(3.75,0)}]
\draw (0, 0) node {\includegraphics[width=0.25\linewidth]{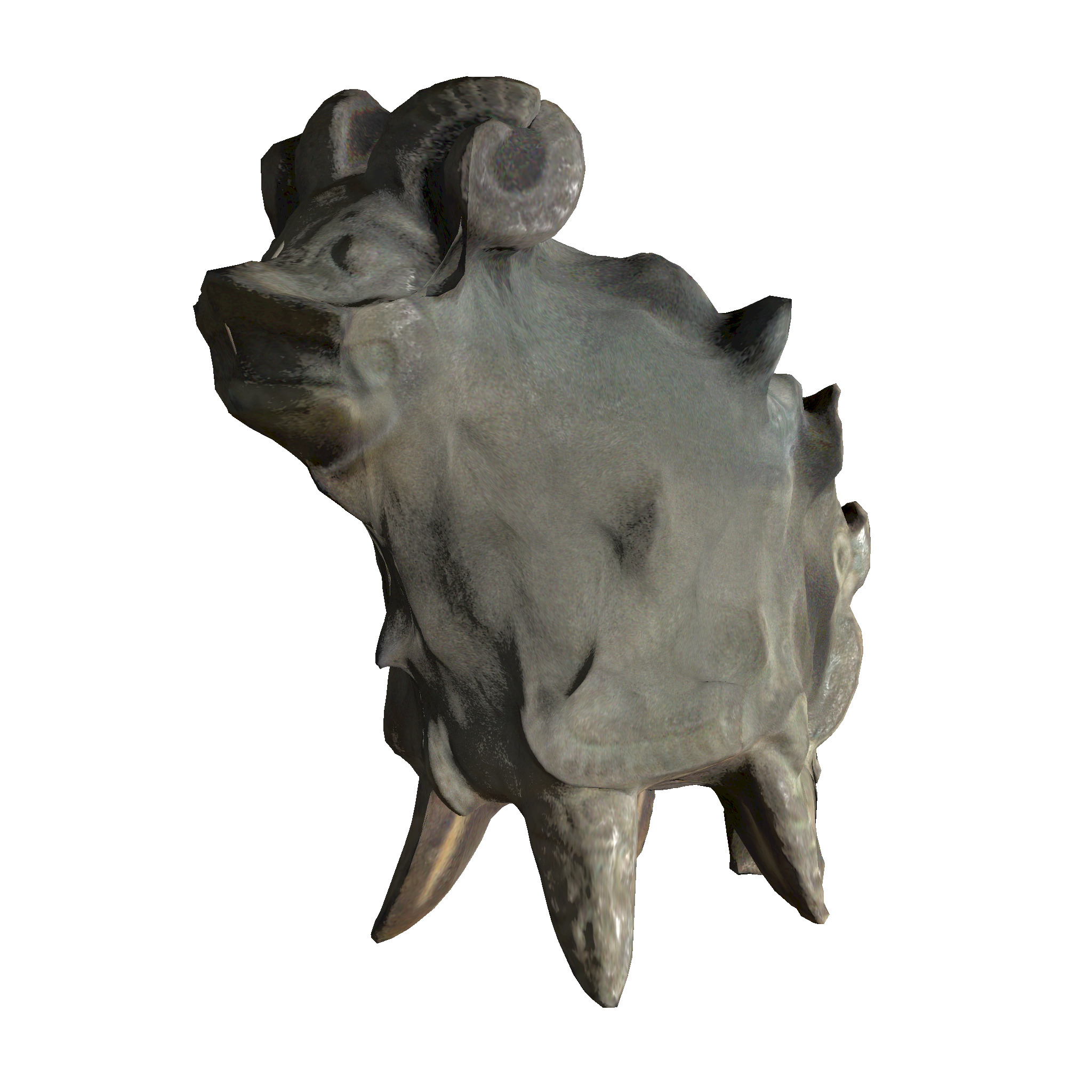} };
\draw (-0.5,-2.5) node {\includegraphics[width=0.15\linewidth]{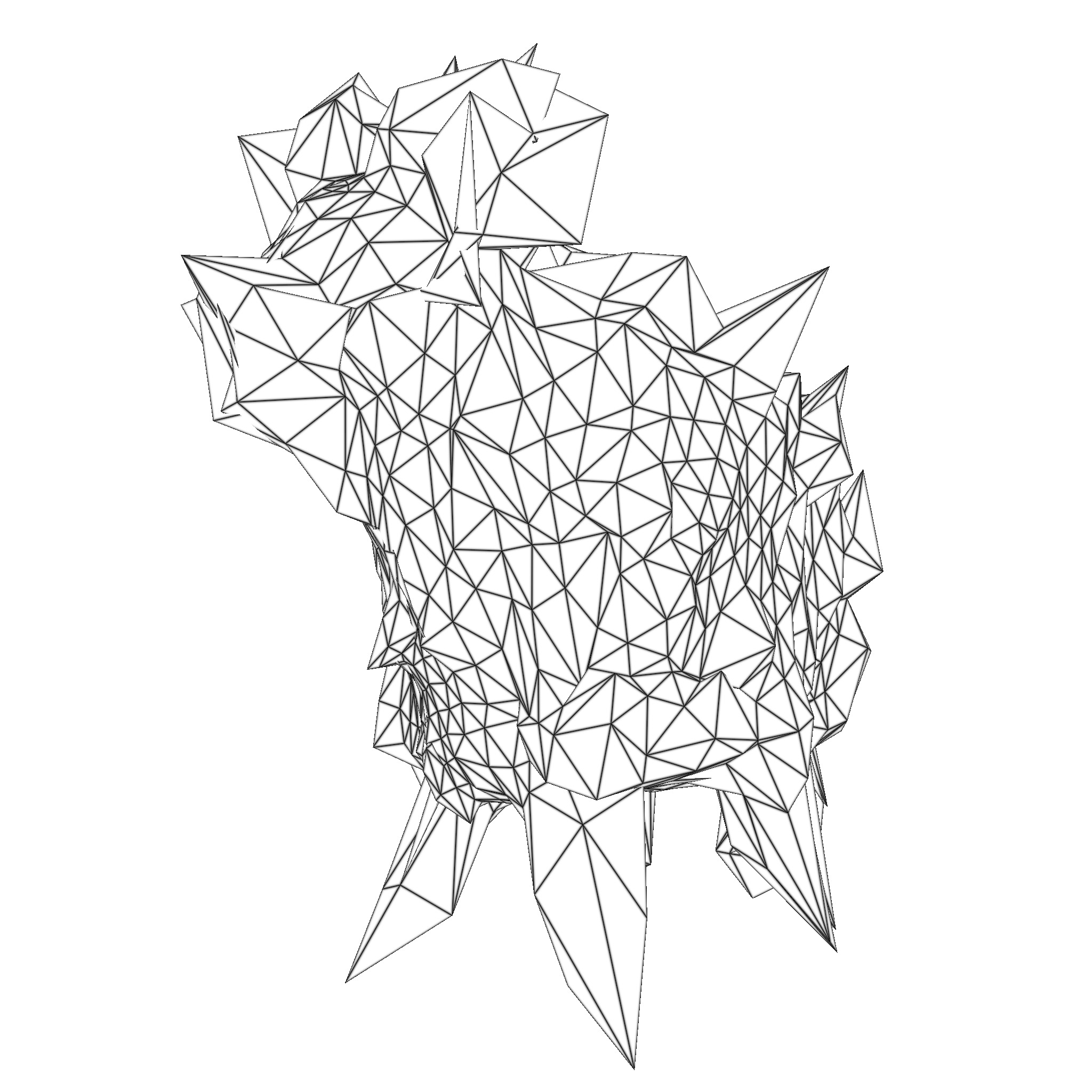} };
\draw (1.1 + 2*0.1, -2.5 - 0*0.1) node {\includegraphics[width=0.1\linewidth, frame]{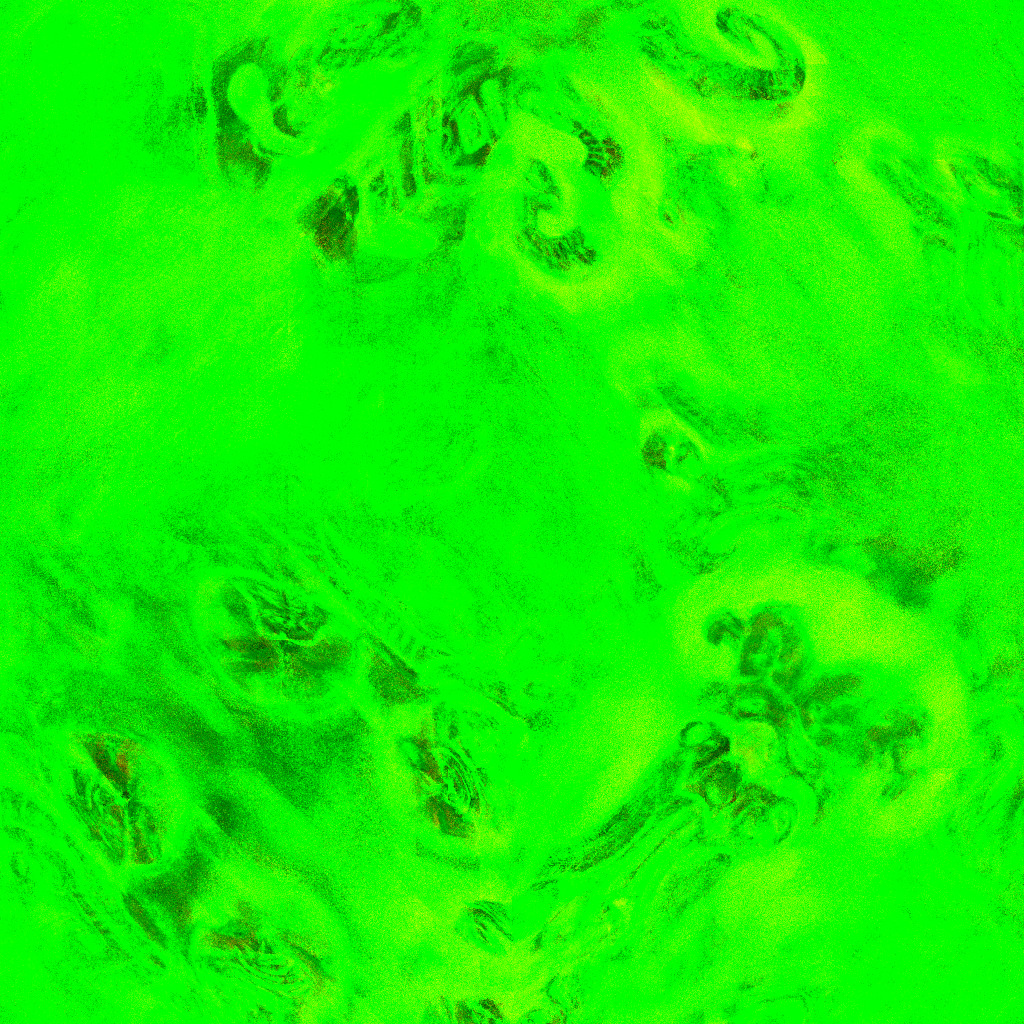}};
\draw (1.1 + 1*0.1, -2.5 - 1*0.1) node {\includegraphics[width=0.1\linewidth, frame]{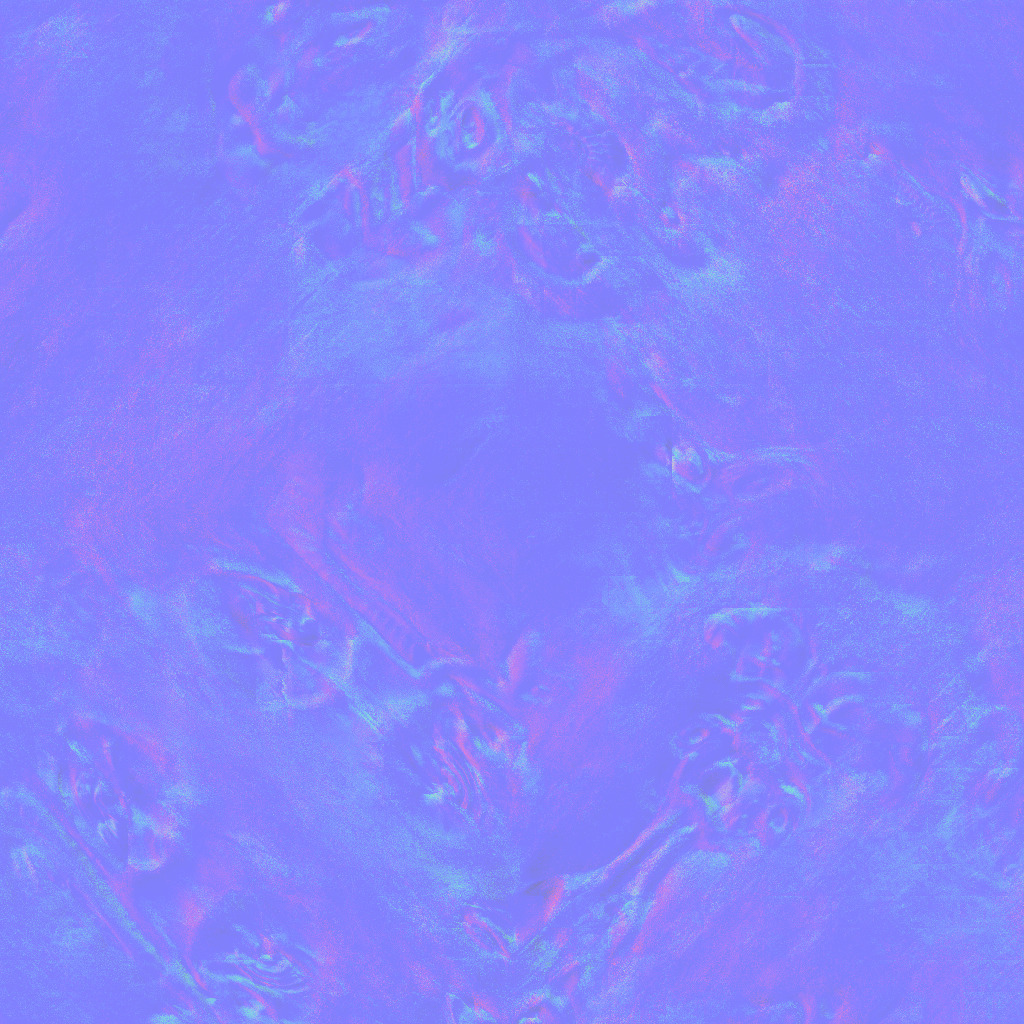} };
\draw (1.1 + 0*0.1, -2.5 - 2*0.1) node {\includegraphics[width=0.1\linewidth, frame]{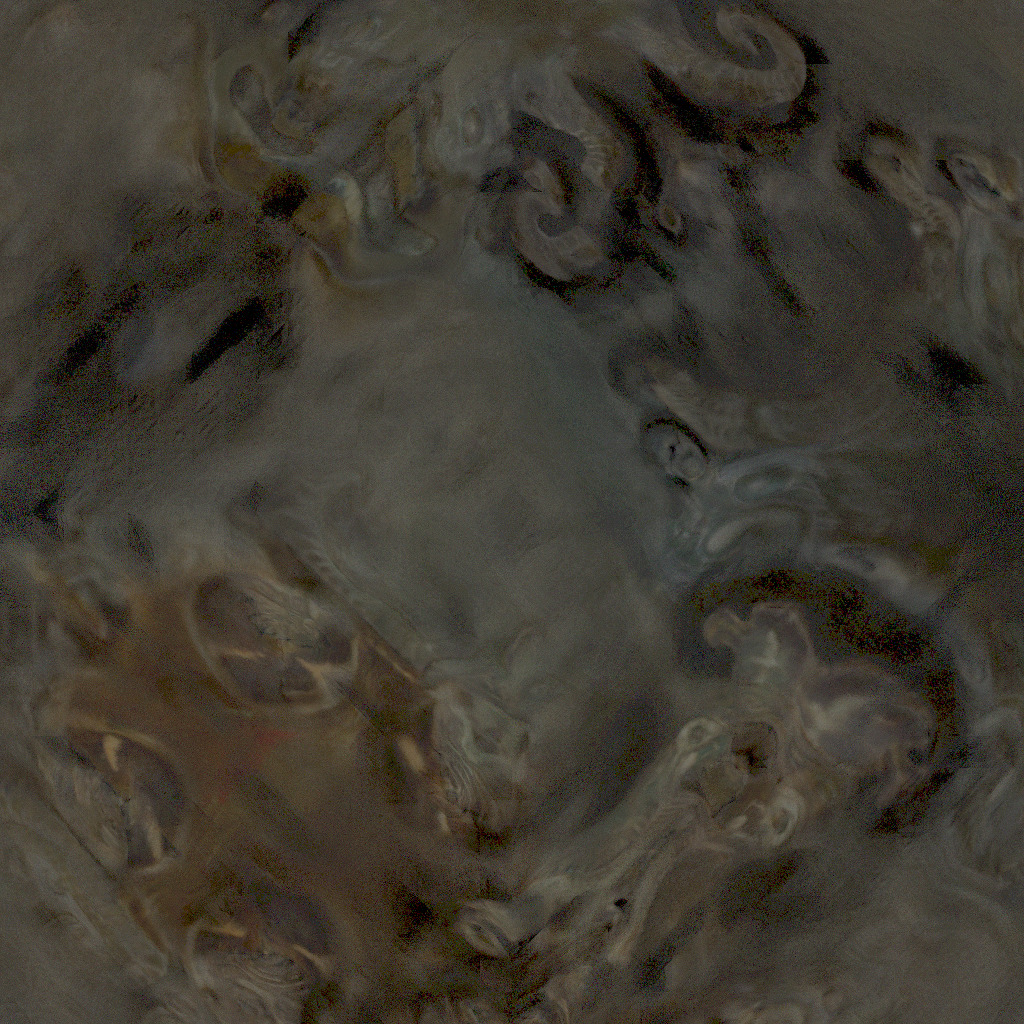}};
\draw (-0.5, -3.6) node {\scriptsize \textbf{control mesh}};
\draw (+1.0 + 0.1, -3.6 - 0.01) node {\scriptsize \textbf{textures}};
\draw (0, 1.75) node {\scriptsize {130 sec}};
\draw (0, 2.05) node {\scriptsize \textbf{our method}};
\end{scope}
\begin{scope}[shift={(3.75*2,0)}]
\draw (0, 0) node {\includegraphics[width=0.25\linewidth]{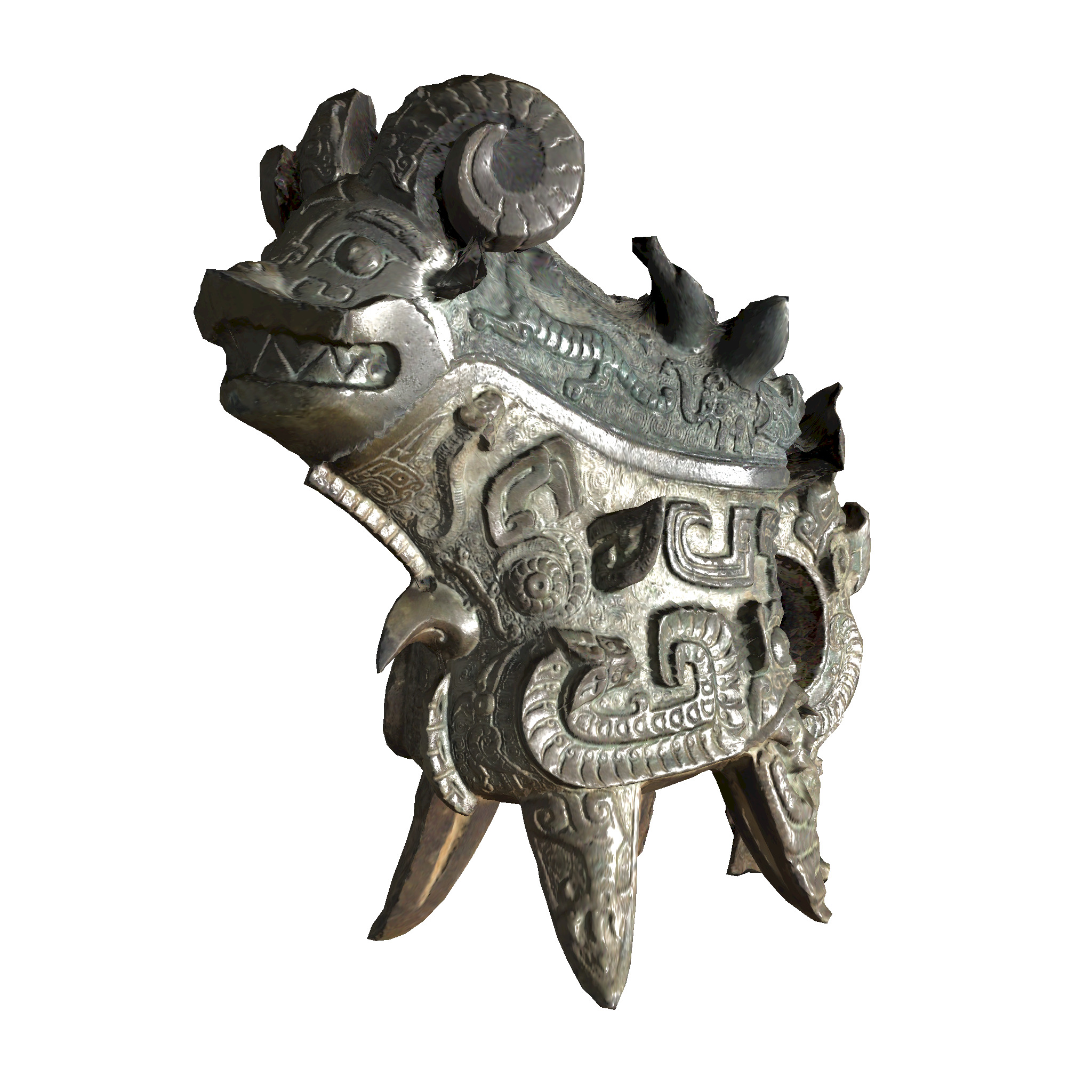} };
\draw (-0.5,-2.5) node {\includegraphics[width=0.15\linewidth]{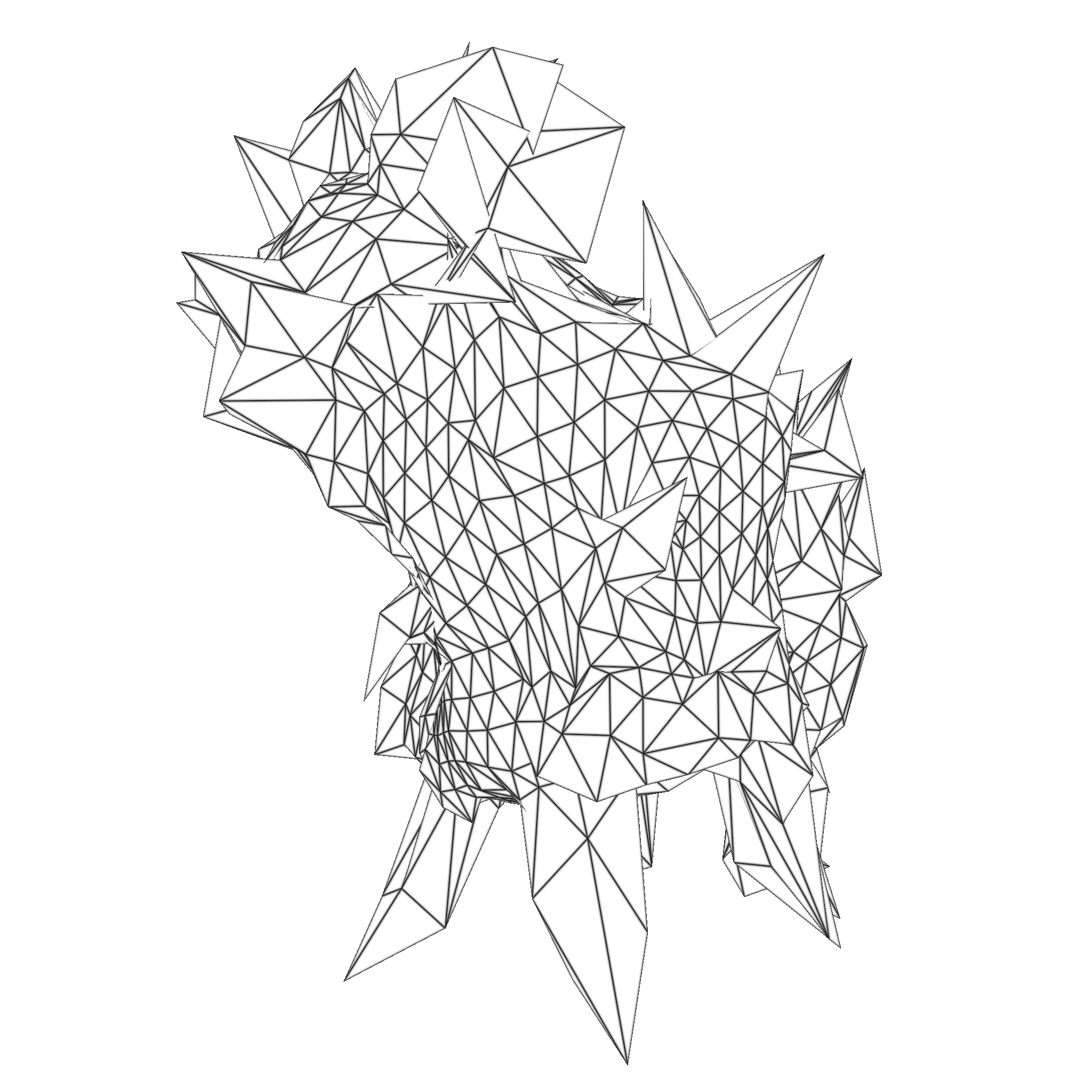} };
\draw (1.1 + 2*0.1, -2.5 - 0*0.1) node {\includegraphics[width=0.1\linewidth, frame]{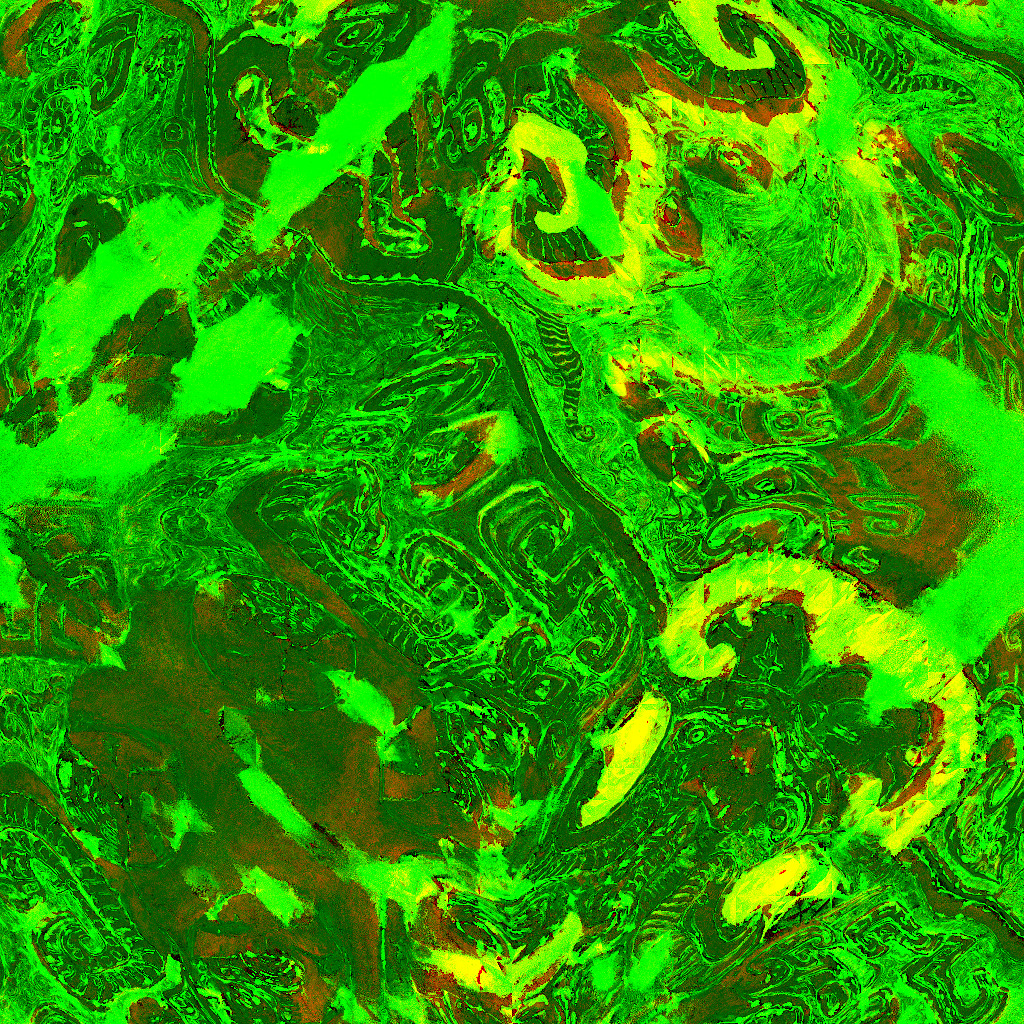}};
\draw (1.1 + 1*0.1, -2.5 - 1*0.1) node {\includegraphics[width=0.1\linewidth, frame]{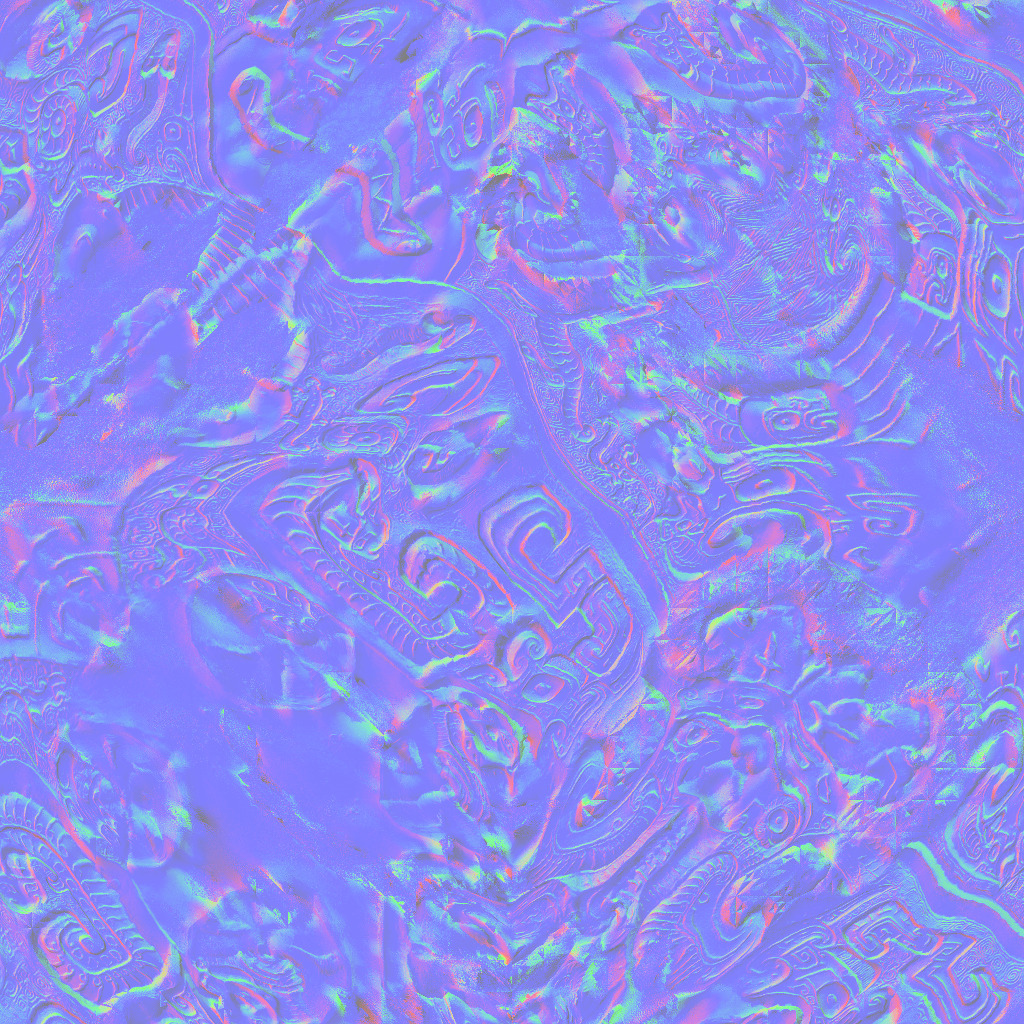}};
\draw (1.1 + 0*0.1, -2.5 - 2*0.1) node {\includegraphics[width=0.1\linewidth, frame]{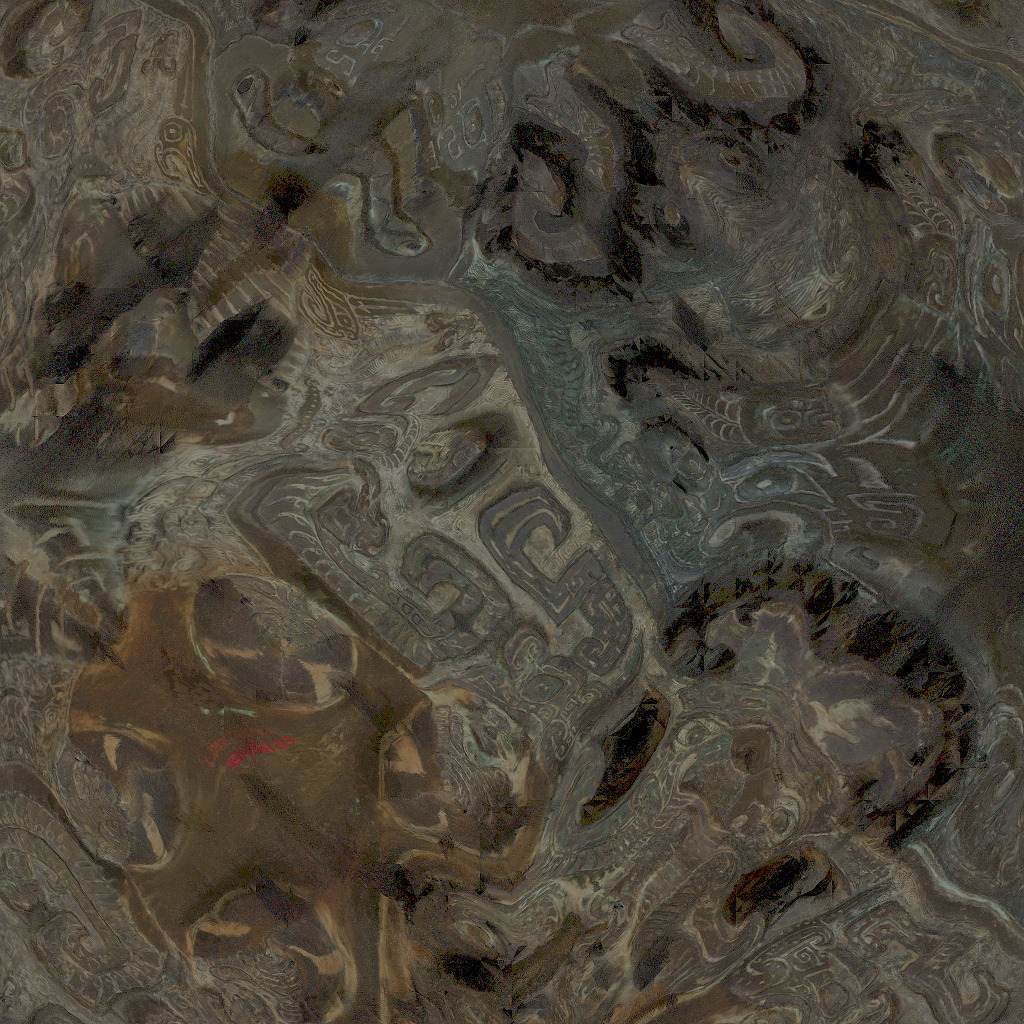}};
\draw (-0.5, -3.6) node {\scriptsize \textbf{control mesh}};
\draw (+1.0 + 0.1, -3.6 - 0.01) node {\scriptsize \textbf{textures}};
\draw (0, 1.75) node {\scriptsize {720 sec}};
\draw (0, 2.05) node {\scriptsize \textbf{our method}};
\end{scope}
\begin{scope}[shift={(3.75*3,0)}]
\draw (0, 0) node {\includegraphics[width=0.25\linewidth]{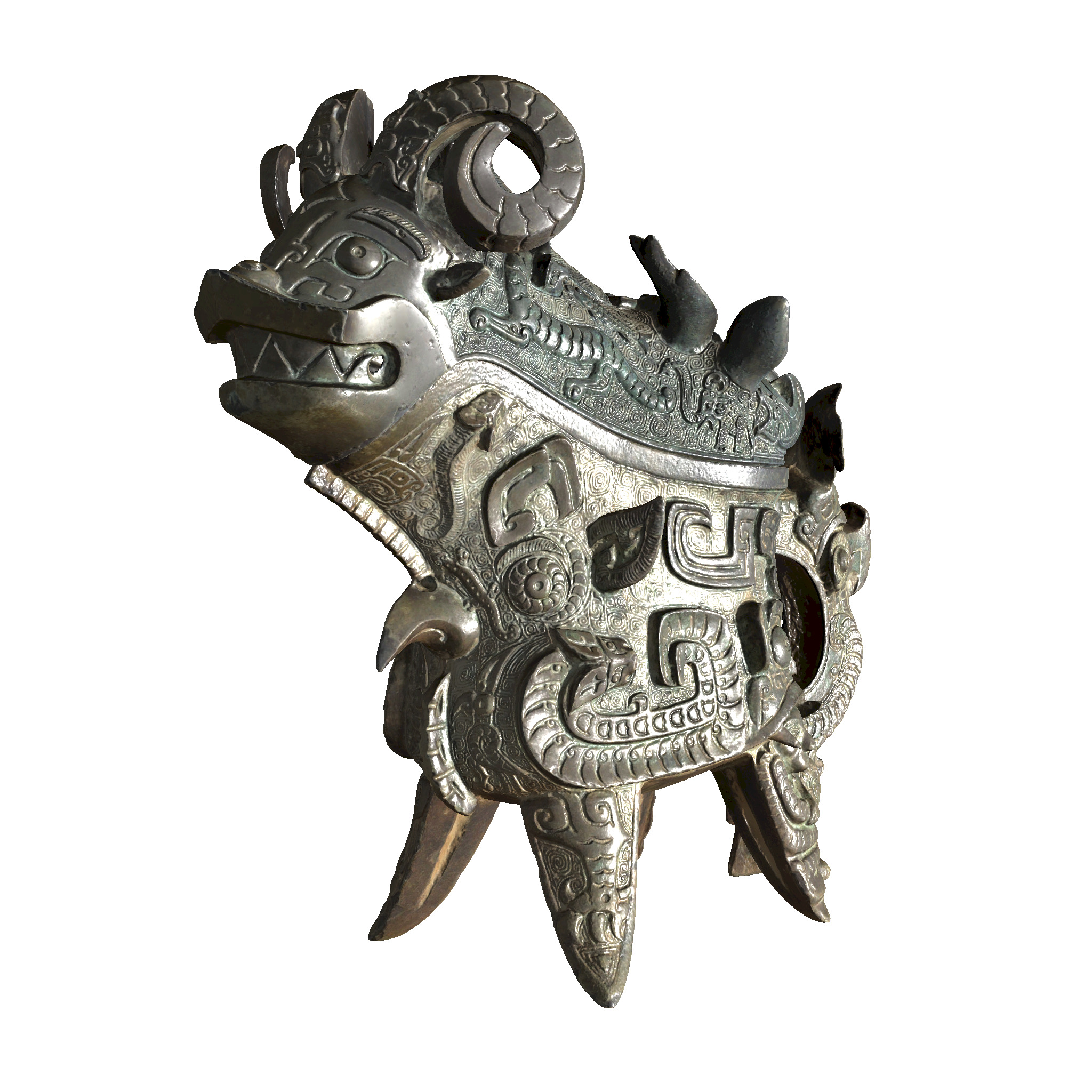}};
\draw (0, 2.05) node {\scriptsize \textbf{reference}};
\end{scope}
\end{tikzpicture}
\end{tabular}
\vspace{-5mm}
  \caption{\label{fig:teaser} 
\textbf{In-engine optimization of assets in their engine-specific geometry and material representations.}
Here, we optimize a control mesh of 2K triangles that controls the tessellation of a Catmull-Clark subdivision surface.
The surface has 50K triangles after two levels of subdivision, which are further displaced, normal mapped and shaded with $1024^2$ physically based textures.
\textbf{Timings are for an Intel Arc 770 GPU}.
}
\end{teaserfigure}


\maketitle

\clearpage
\section{Introduction}

\paragraph{Motivation for differentiable rendering.}

A differentiable renderer is a rendering engine that computes a 2D image for a given 3D scene and has, in addition, the ability to provide gradients for the 3D scene parameters via backpropagation through the rendering calculations.
The benefits of having these gradients is that it makes possible to optimize the 3D scene parameters to obtain a target 2D image via gradient descent.
This allows for many applications such as object placement~\cite{rhodin2015versatile}, object reconstruction~\cite{kato2019learning,wu2023unsupervised}, model simplification~\cite{hasselgren2021appearance}, material estimation~\cite{azinovic2019inverse}, etc.

\paragraph{Objective.}

We assume that a rasterization engine is available and we wish to use differentiable rendering to optimize assets for their final in-engine rendering.
Ideally, the solution should keep the workflow simple and self-contained, \textit{i.e.} without using other tools and dependencies than the engine itself. 
In this context, implementing a renderer from scratch within a differentiable frameworks such as Dr.JIT~\cite{jakob2020drjit} or Slang.D~\cite{bangaru2023slangd} is not an option. 
Using existing differentiable rasterizers such as \textsc{nvDiffRast}~\cite{laine2020diffrast} requires externalizing the workflow and relying on external (sometimes vendor-specific) dependencies, which is also problematic. 
This is why we aim at transforming an existing non-differentiable rasterizer into a differentiable one.

\paragraph{Contribution.}

Our method is based on the concept of \textit{Stochastic Gradient Estimation}~\cite{fu2005}, a stochastic variant of finite differentiation that allows for estimating gradients without a differentiable framework.
However, akin to finite differentiation, this method does not scale to high-dimensional problems: the more dimensions, the noisier the gradient estimates, the more optimization steps are required.
Our idea is to cut down the dimensionality by estimating gradients on a per-pixel basis rather than on the whole image. 
Indeed, the number of parameter contributing to a given rasterized pixel is of tractable dimensionality, regardless of the total number of parameters in the scene. 
This idea yields a method to make an existing rasterizer differentiable. Namely:\\
\vspace{-5mm}\\
\begin{itemize}
\item It is \textbf{simple to implement}. Our base differential rasterization component consists of adding ID/UV-buffers to the existing raster targets and two compute shaders. 
\vspace{-2mm}\\

\item It \textbf{keeps the workflow self-contained} by bringing the benefits of differentiable rasterization to an existing conventional rasterizer without requiring external dependencies.
\vspace{-2mm}\\

\item It is \textbf{cross-platform} since it uses only conventional graphics API functionalities. 
This is a significant bonus point for adoption given that existing differential rendering solutions are bound to vendor-specific hardware and/or software.
\vspace{-2mm}\\

\item It is \textbf{efficient and scales well in scene complexity}. 
We optimize scenes with 1M+ parameters on a customer GPU. 
Furthermore, despite stochastic differentiation with noisy gradients being theoretically less efficient than backpropagated differentiation with clean gradients, our implementation is qualitatively on par with \textsc{nvDiffRast}~\cite{laine2020diffrast} in our experiments. 
This is because the gain in speed of remaining in an existing and well-optimized rasterization engine, in contrast to switching to a significantly slower Pytorch environment, finally compensates for the slower convergence due to noisier gradients. 
\vspace{-2mm}\\

\item It \textbf{covers multiple use cases}. We estimate gradients for meshes, displacement mapping, Catmull-Clark subdivision surfaces~\cite{CatmullClark78}, semi-transparent geometry, physically based materials, 3D volumetric data and 3D Gaussian Splats~\cite{kerbl2023}.
\vspace{-2mm}\\

\item Our scope is \textbf{raster graphics} (direct visibility only). Our method does not cover further rendering events such as shadows or multiple-bounce illumination. 
\end{itemize}

\clearpage
\section{Previous Work}

\noindent \phantom{a} \vspace{-5mm}
\subsection{Differentiable rasterization}
\label{sec:previous_diff_rast}

Differentiable rasterization usually revolves around smoothing the discontinuous visibility function to make it differentiable~\cite{loper2014opendr,kato2018neural,liu2019soft}.
The state-of-the-art framework is \textsc{nvDiffRast}~\cite{laine2020diffrast}, a performant and modular differentiable rasterizer, which we use as a comparison baseline. 
\\
    
\noindent 
Note that all these methods require a Pytorch/Tensorflow context with external dependencies and are often vendor-specific. 
We position our method as an in-engine, dependency-free and cross-platform alternative.
Our experiments on mesh and texture optimization show qualitatively that it is competitive in terms of optimization speed for these applications. 

\subsection{Stochastic Gradient Estimation}

The concept of estimating gradients in a stochastic manner by applying random perturbations to the input comes in many flavors and under many names such as \textit{Stochastic Gradient Estimation}~\cite{fu2005}, \textit{Monte Carlo Gradient Estimation}~\cite{patelli2010monte}, \textit{Gradient Estimation Via Perturbation Analysis}~\cite{glasserman1991}, \textit{Perturbed Optimization}~\cite{berthet2020}, and many others. \\
    
\noindent
We use one of the variants presented in \emph{Stochastic Gradient Estimation}~\cite{fu2005}, a stochastic variant of finite differentiation. 
We found it to be the simplest one to convey while proving competitive enough in our experiments.

\subsection{Differentiable Rendering with Stochastic Gradient Estimation}

Variants of stochastic gradient estimation have already been transposed to the field of rendering~\cite{lelidec2021differentiable,fischer2023plateau}.
In this context, it consists of randomly perturbing the 3D scene parameters such that the variation of the 2D image error averaged over the perturbations provides an unbiased estimate of the 3D scene parameter gradients. 
With this, the scene parameters can be optimized to match a target image. 
The approach of Fischer and Ritschel.~\shortcite{fischer2023plateau} is especially close to ours because it can be directly implemented within an existing renderer without further dependencies. 
However, these methods do not scale to high-dimensional problems: the more dimensions, the noisier the gradient estimates, the more optimization steps are required.
They are thus limited to low-dimensional problems such as 6D pose estimation or optimizing low-poly meshes. 
\\
    
\noindent
The key difference of our method is that it estimates gradients on a per-pixel basis rather than on the whole image. 
Thanks to this, it scales up to scenes with 1M+ parameters such as dense or textured meshes.

\subsection{Differentiable Monte Carlo Rendering}

Differentiable Monte Carlo path tracers that account for illumination effects beyond direct visibility have been developed~\cite{li2018redner,david2019mitsuba,zhang2020path,vicini2021path}.
They come with dedicated algorithms to cover difficult cases such as silhouettes and shadows~\cite{loubet2019reparameterizing,bangaru2020warpedsampling,yan2022efficient}.\\
    
\noindent Our method is solely based on rasterization (direct visibility) and excludes multiple-bounce effects.
We hence do not compete with this line of work.

\clearpage
\clearpage
\section{Background on Stochastic Gradient Estimation}

In this section, we provide background on \textit{Stochastic Gradient Estimation}, a stochastic variant of the finite-difference method. 
We refer the reader to the work of Fu~\shortcite{fu2005} for more details. 

\paragraph*{Problem statement.}

We consider a $d$-dimensional space of parameters $\boldsymbol\theta = (\theta_1, .., \theta_d)$ where $d$ is large and an objective function $f(\boldsymbol\theta) \in \mathbb{R}^+$.
Our objective is to solve the minimization problem:
\begin{align}
\min_{\boldsymbol\theta \in \mathbb{R}^d} \hspace{3mm} f(\boldsymbol\theta).
\end{align}
For this purpose, we wish to use a gradient-descent optimizer. 
We thus need a way to evaluate
\begin{align}
\frac{\partial f}{\partial \boldsymbol\theta}~=~~?
\end{align}
In machine-learning frameworks (Pytorch/Tensorflow), this gradient is estimated via backpropagation. 
We wish to find an alternative way to estimate this gradient when a backpropagation machinery is not available. 

\paragraph*{Finite difference.}

The classic finite-difference method method computes a numerical derivative in each dimension by perturbing each component with a small offset:
\begin{align}
\label{eq:finite_difference}
\frac{\partial f}{\partial \theta_i}
\hspace{3mm}
=
\hspace{3mm}
\frac{f(\boldsymbol\theta + \boldsymbol b_i \odot \boldsymbol\epsilon) - f(\boldsymbol\theta - \boldsymbol b_i \odot \boldsymbol\epsilon)}{2 \, \epsilon_i}.
\end{align}
where $\boldsymbol\epsilon = \left(\epsilon_1, ..,\epsilon_d\right)$ is a user-defined perturbation magnitude vector and $b_i = (0..0, 1, 0..0)$ is the $i$-th basis vector. 
We note $\boldsymbol b_i \odot \boldsymbol\epsilon$ the element-wise product of both vectors.
The limitation of this approach is that it requires two evaluations of $f()$ per dimension, which makes it untractable in high-dimensional spaces.

\paragraph*{Stochastic finite difference.}

To overcome the dimensionality problem of the finite-difference method, a variant consists of randomly perturbing all the dimensions simultaneously to obtain a stochastic estimator of the gradient:
\begin{align}
\label{eq:stochastic_finite_difference}
\widehat{
\frac{\partial f}{\partial \theta_i}
}
\hspace{3mm}
=
\hspace{3mm}
\frac{f(\boldsymbol\theta + \boldsymbol s \odot \boldsymbol\epsilon) - f(\boldsymbol\theta - \boldsymbol s \odot \boldsymbol\epsilon)}{2 \, s_i \, \epsilon_i}.
\end{align}
where $\boldsymbol s = \left(s_1, .., s_d\right)$ is a random sign vector that contains independent variables $s_i \in \{-1,+1\}$ where each sign has probability $\frac{1}{2}$.
The advantage of this method is that two evaluations of $f()$ yield an estimation of the gradient regardless of the number of dimensions. 
The downside is that the estimator is stochastic, \text{i.e.} it is a random variate that is correct on expectation\footnote{
Note that finite-difference methods are biased.
The expectation of Equation~(\ref{eq:stochastic_finite_difference}) is thus an approximation of the exact gradient, depending on the perturbation magnitude $\boldsymbol\epsilon$. 
We explain how to set this parameter in practice in Section~\ref{sec:diffast}.} but exhibits some variance.  
Furthermore, the more dimensions, the higher the variance of the estimator is. 
In summary, replacing Equation~(\ref{eq:finite_difference}) by Equation~(\ref{eq:stochastic_finite_difference}) means trading accuracy for performance. 
\section{Differentiable Rasterization with Stochastic Gradient Estimation}
\label{sec:diffast}

We now apply \textit{Stochastic Gradient Estimation} to differential rasterization, where the objective is to optimize a 3D scene such that a rasterized 2D image produced with this scene matches a target image. 
To do this, we need to estimate the gradients of the rasterization computations. 

\subsection{Notations}

In this context, the vector $\boldsymbol\theta \in \mathbb{R}^d$ represents a 3D scene defined by a set of parameters, typically geometry, textures, etc. 
A rasterizer computes a 2D image $I(\boldsymbol\theta)$ using this 3D scene. 
Finally, the objective function $f(\boldsymbol\theta) = \|I(\boldsymbol\theta) - I\|^2$ is the error between the rasterized image $I(\boldsymbol\theta)$ and a target image $I$.
We summarize these notations in Table~\ref{tab:notation}.

\begin{table}[!h]
\begin{tabular}{@{} c @{\hspace{10mm}} c @{\hspace{10mm}} l @{}}
Name & Domain & Description \\
\toprule
$\boldsymbol\theta = (\theta_1, .., \theta_d)$ & $\mathbb{R}^d$ & 3D scene parameters (geometry, textures, etc.) \\
$\boldsymbol s = (s_1, .., s_d)$ & $ \{-1,+1\}^d$ & random sign vector \\
$\boldsymbol \epsilon = (\epsilon_1, .., \epsilon_d)$ & $ \mathbb{R}^d$ & perturbation magnitude vector \\
$I$ & $ \mathbb{R}^{3\times W \times H}$ & 2D target RGB image \\
$I(\boldsymbol\theta)$ & $ \mathbb{R}^{3\times W \times H}$ & 2D rasterized RGB image \\
$I_{w,h}$ & $ \mathbb{R}^{3}$ & pixel $(w,h)$ in target image $I$\\
$I_{w,h}(\boldsymbol\theta)$ & $ \mathbb{R}^{3}$ & pixel $(w,h)$ in rasterized image $I(\boldsymbol\theta)$ \\
$f(\boldsymbol\theta) = \|I(\boldsymbol\theta) - I\|^2$ & $ \mathbb{R}^+$ & $l_2$ image error \\
$f_{w,h}(\boldsymbol\theta)=\|I_{w,h}(\boldsymbol\theta) - I_{w,h}\|^2$ & $\mathbb{R}^+$ & $l_2$ pixel $(w,h)$ error \\
\bottomrule 
\end{tabular}
\vspace{0mm}
\caption{\label{tab:notation}Notations.}
\vspace{-10mm}
\end{table}

\subsection{Per-Pixel Formulation}
\label{sec:per_pixel_formulation}

\paragraph{Motivation}

As explained previously, the downside of Equation~(\ref{eq:stochastic_finite_difference}), is that the stochastic gradient estimate is noisy, especially in a high-dimensional parameter space.
Intuitively, in our rasterization use case, a large part of this noise can be explained by the fact that the error over the whole image (the error in every pixel) contributes to all the scene parameters.
Even if a parameter is never used to compute a pixel it receives noisy gradients from this pixel.  
In theory, this is not a problem because the noisy gradients conveyed by a pixel not impacted by a parameter are null on expectation. 
However, in practice, the noisy gradients dramatically burden the gradient decent. 
In Section~\ref{sec:results}, we show that this method can hardly be used as is to optimize large scenes. 
We thus propose a per-pixel gradient computation approach that alleviates this problem and makes the method usable.

\paragraph{Derivation}

The error we use is the $l_2$ error, which is sum of per-pixel errors:
\begin{align}
f(\boldsymbol\theta) 
= \sum_{(w,h) \in W\times H} f_{w,h}(\boldsymbol\theta),
\end{align}
and the gradient can be defined in the same way:
\begin{align}
\frac{\partial f}{\partial \theta_i} 
= \sum_{(w,h) \in W\times H} \frac{\partial f_{w,h}}{\partial \theta_i}.
\end{align}
Note that if the parameter $\theta_i$ is not implicated in the computation of pixel $(w,h)$ then $\frac{\partial f_{w,h}}{\partial \theta_i} = 0$.
We can thus rewrite the gradient with a sparse sum where only impacted pixels contribute:
\begin{align}
\label{eq:gradient_per_pixel}
\frac{\partial f}{\partial \theta_i} 
= \sum_{(w,h) \text{ {\tiny impacted by }} \theta_i} \frac{\partial f_{w,h}}{\partial \theta_i}.
\end{align}
By applying the estimator of Equation~(\ref{eq:stochastic_finite_difference}) to Equation~(\ref{eq:gradient_per_pixel}) we obtain the stochastic gradient estimate our method is based on:
\begin{align}
\label{eq:stochastic_finite_difference_pixel}
\widehat{
\frac{\partial f}{\partial \theta_i}
}
\hspace{3mm}
=
\hspace{3mm}
\sum_{(w,h) \text{ {\tiny impacted by }} \theta_i}
\frac{f_{w,h}(\boldsymbol\theta + \boldsymbol s \odot \boldsymbol\epsilon) - f_{w,h}(\boldsymbol\theta -\boldsymbol s \odot \boldsymbol\epsilon)}{2 \, s_i \, \epsilon_i}.
\end{align}
In Section~\ref{sec:overview}, we show how to implement this equation with a rasterizer and compute shaders.

\subsection{Overview}
\label{sec:overview}

\begin{figure}[!h]
\begin{center}
\scalebox{1.15}{
\input{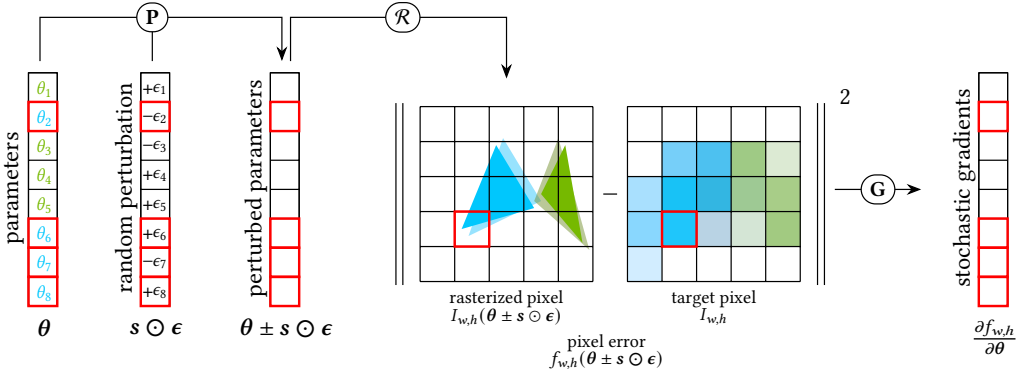}
}
\end{center}
\caption{\label{fig:overview} 
\textbf{Overview of our differentiable rasterizer.}
The first compute shader (\textbf{P}) perturbs the scene parameters before they are rasterized (${\mathcal{R}}$).
The second compute shader (\textbf{G}) accumulates the error differences, which provide a gradient estimate. 
The key point of our approach is that it accumulates the contribution of a pixel (in red in the images) only in its contributing parameters (in red in the vectors). 
}
\end{figure}

Our differentiable rasterizer implements Equation~(\ref{eq:stochastic_finite_difference_pixel}) with 2 compute shaders \textbf{P} (perturbation) and \textbf{G} (gradient) in addition to the rasterizer $\mathcal{R}$.
We provide an overview of our pipeline in Figure~\ref{fig:overview}. 
~\vspace{2mm}\\

~\vspace{-10mm}\\

\begin{algorithm}
\centering
\caption{Compute shader \textbf{P} (perturbation)}\label{algo:P}
\begin{algorithmic}
\Require thread ID $i$
\State \textbf{load} $\theta_i$, $\epsilon_i$ \Comment{load 2 float}
\State $s_i$ = \textbf{randomsign}() \Comment{hash function \cite{jarzynski2020hash}}
\State \textbf{store} $s_i\epsilon_i$, $\theta_i+s_i\epsilon_i$, $\theta_i-s_i\epsilon_i$ \Comment{store 3 float}
\end{algorithmic}
\end{algorithm}

~\vspace{-10mm}\\

\begin{algorithm}
\centering
\caption{Compute shader \textbf{G} (gradient)}\label{algo:G}
\begin{algorithmic}
\Require thread ID $(w,h)$
\State \textbf{load} 
$I_{w,h}(\boldsymbol\theta+\boldsymbol s\odot\boldsymbol\epsilon)$, 
$I_{w,h}(\boldsymbol\theta-\boldsymbol s\odot\boldsymbol\epsilon)$, 
$I_{w,h}$ \Comment{load 3 float3 (3$\times$ rgb)}
\State $f_{w,h}(\boldsymbol\theta+\boldsymbol s\odot\boldsymbol\epsilon) = \left\|I_{w,h} - I_{w,h}(\boldsymbol\theta+\boldsymbol s\odot\boldsymbol\epsilon) \right\|^2$ 
\State $f_{w,h}(\boldsymbol\theta-\boldsymbol s\odot\boldsymbol\epsilon) = \left\|I_{w,h} - I_{w,h}(\boldsymbol\theta-\boldsymbol s\odot\boldsymbol\epsilon) \right\|^2$ 
\For{each parameter $\theta_i$ contributing to pixel $(w,h)$} \Comment{implementation of Equation~(\ref{eq:stochastic_finite_difference_pixel})}
\State \textbf{load} $s_i\epsilon_i$ \Comment{load 1 float}
\State \textbf{AtomicAdd}$\left(\frac{\partial f}{\partial \theta_i} \leftarrow 
\frac{f_{w,h}(\boldsymbol\theta + \boldsymbol s \odot \boldsymbol\epsilon) - f_{w,h}(\boldsymbol\theta -\boldsymbol s \odot \boldsymbol\epsilon)}{2 \, s_i \, \epsilon_i}\right)$  \Comment{atomic add 1 float}
\EndFor
\end{algorithmic}
\end{algorithm}

\paragraph*{User-defined perturbation magnitude vector $\boldsymbol \epsilon$}

Our general methodology to set the perturbation magnitude vector $\boldsymbol \epsilon$ is that the perturbation should produce a small but measurable change in the rasterized image.
If $\theta_i$ is a triangle vertex coordinate, we set $\epsilon_i$ such that it results in a perturbation of 1-2 pixels on average in screen-space. 
If $\theta_i$ is a texel (or voxel) parameter, we set $\epsilon_i$ to the quantization of the texture (or volume) data format.

\clearpage

\paragraph*{Compute shader \textbf{P} (perturbation)}

We launch this compute shader over $d$ threads (the number of scene parameters) that execute Algorithm~\ref{algo:P}.
The shader computes the perturbed scene parameters $\boldsymbol\theta+\boldsymbol s \odot \boldsymbol\epsilon$ and $\boldsymbol\theta-\boldsymbol s \odot \boldsymbol\epsilon$.
Its main ingredient is the generation of the random sign vector $\boldsymbol s$ via $\textbf{randomsign}()$, which we implement with a random hash function~\cite{jarzynski2020hash}.

\paragraph*{Rasterization $\mathcal{R}$}

We rasterize the scenes of parameters $\boldsymbol\theta+\boldsymbol s \odot \boldsymbol\epsilon$ and $\boldsymbol\theta-\boldsymbol s \odot \boldsymbol\epsilon$ and obtain two images $I(\boldsymbol\theta+\boldsymbol s\odot\boldsymbol\epsilon)$ and $I(\boldsymbol\theta-\boldsymbol s\odot\boldsymbol\epsilon)$.

\paragraph*{Compute shader \textbf{G} (gradient)}

We launch this compute shader over $W \times H$ threads (the number of pixels) that execute Algorithm~\ref{algo:G}.
The shader computes the pixel errors $f_{w,h}(\boldsymbol\theta+\boldsymbol s\odot\boldsymbol\epsilon)$ and $f_{w,h}(\boldsymbol\theta-\boldsymbol s\odot\boldsymbol\epsilon)$ between the perturbed-scene images $I(\boldsymbol\theta+\boldsymbol s\odot\boldsymbol\epsilon)$ and $I(\boldsymbol\theta-\boldsymbol s\odot\boldsymbol\epsilon)$ and the target image $I$.
Once these errors are available, they provide the gradient estimate for each parameter $i$ contributing to pixel $(w,h)$ following Equation~(\ref{eq:stochastic_finite_difference_pixel}).
We add the result to the gradient estimate using an \textbf{AtomicAdd} operation to avoid interferences between multiple threads (pixels) adding simultaneously their gradient contribution to the same parameter.
Note that the critical point of this algorithm is the ability to loop over each parameter $i$ contributing to pixel $(w,h)$. 
We explain how we achieve this in practice for each type of primitive in Section~\ref{sec:primitives}.

\subsection{Primitives Implementation}
\label{sec:primitives}

Our method uses different strategies depending on the type of content being optimized. 
For each kind of primitive, we explain how to implement the loop in compute shader \textbf{G} (Algorithm~\ref{algo:G}) over the parameters $\theta_i$ contributing to a given pixel $(w, h)$. 

\paragraph{Opaque geometry} 

We represent opaque geometry with triangles meshes defined by a vertex buffer that stores the 3D vertices and an index buffer that stores the vertices of each triangle. 
We modify the rasterization pass $\mathcal{R}$ such that, in addition to the RGB output, it rasterizes an ID buffer that contains the index of the rasterized triangle in each pixel.
In the compute shader \textbf{G}, we sample the ID buffer for each pixel $(w, h)$ to identify the triangle seen by this pixel and use the index buffer to recover the vertices of this triangle.

\paragraph{Transparent geometry} 

In the case of transparent geometry, we further modify our rasterization pass $\mathcal{R}$ to support transparent front-to-back rendering with a pre-sorting pass, and output a deep ID buffer with multiple triangle IDs per pixel. 
This gives us an ordered list of the triangles seen by a pixel. 
We go through this list in compute shader \textbf{G} and proceed in a similar manner as described above for each triangle in the list.

\paragraph{Textures} 

To optimize texture content, we further modify the rasterization pass $\mathcal{R}$ to rasterize a UV buffer in addition to the RGB output and the ID buffer.
It contains the UV coordinates used to fetch the texture in each pixel.
In the compute shader \textbf{G}, we use these UV coordindates to recover the texel that contributed to the pixel. 
Note that, in theory, a pixel should contribute to the gradient estimates of all the texels that fall within its texture-space elliptical footprint.
In practice, we find that doing so only for the texel closest to the center of the footprint is sufficient if the rendering resolution is high enough to avoid sub-pixel scale texels.

\paragraph{Volumes} 

To render volumetric content, we ray-march a 3D texture during the rasterization pass $\mathcal{R}$.
In the compute shader \textbf{G}, we implement the loop as another pass of ray-marching where each encountered voxel receives gradient update.

\clearpage
\clearpage

\section{Application to 3D scene Optimization}
\label{sec:application}

We explain how to use the differential rasterizer described in Section~\ref{sec:diffast} to optimize 3D scenes.

\paragraph*{Gradient accumulation loop}

The differential rasterizer introduced in Section~\ref{sec:diffast} evaluates Equation~(\ref{eq:stochastic_finite_difference_pixel}) to obtain a stochastic (noisy) estimate of the gradient. 
The noisiness of these gradients can burden the gradient descent. 
It is possible to obtain a lower-variance estimator by averaging $N$ stochastic gradient estimates:
\begin{align}
\label{eq:stochastic_finite_difference_pixel_N}
\widehat{
\frac{\partial f}{\partial \theta_i}
}
\hspace{3mm}
=
\hspace{3mm}
\frac{1}{N}
\sum_{n=1}^N
\sum_{(w,h) \text{ {\tiny impacted by }} \theta_i}
\frac{f_{w,h}(\boldsymbol\theta + \boldsymbol s^{(n)} \boldsymbol\epsilon) - f_{w,h}(\boldsymbol\theta -\boldsymbol s^{(n)} \boldsymbol\epsilon)}{2 \, s^{(n)}_i \, \epsilon_i}.
\end{align}
where the $n$th estimation uses a different random sign vector $s^{(n)}$. 
We implement this as a loop that repeats $N$ times the steps \textbf{P}, $\mathcal{G}$, and \textbf{G}.
Note this averaging loop is usually necessary anyways even with deterministic differential rasterizers because there are other sources of noise in the gradients such as the random choice of the point of view.
In our case, we randomize our sign vector $\boldsymbol s^{(n)}$ simultaneously with these other random variables in each iteration $n$.

\paragraph*{Gradient-descent optimizer}

After the gradient accumulation loop, we use the gradient estimate to make a gradient descent over the parameters $\boldsymbol \theta$.
Since our gradient estimate is stochastic, it is preferable to use a gradient-descent optimizer specifically designed for performing stochastic gradient descent such as Adam~\cite{kingma2015adam}.
We implement Adam in a compute shader launched over $d$ threads (the number of scene parameters) that takes $\boldsymbol \theta$ and $\frac{\partial f}{\partial \boldsymbol \theta}$ as inputs and updates $\boldsymbol \theta$.
We use it with its default parameters $\beta_1=0.9$ and $\beta_2=0.999$ and we set the learning rate of each parameter $\theta_i$ to the same value as its perturbation amplitude $\epsilon_i$ in all our experiments.
Note that Adam is invariant to constant scaling factors. In our implementation, we do not perform the division by the 
constants $2$, $\epsilon_i$ and $N$ in the denominators of Equation~(\ref{eq:stochastic_finite_difference_pixel_N}).

\paragraph*{Additional non-gradient-based optimizations}

A gradient descent remains a local exploration of the optimization landscape.
In some cases, even with good gradient estimates, the gradient-descent optimizer might be stuck in local minima. 
Some applications require additional non-gradient-based optimization to converge successfully.
For instance, the triangles in Figure~\ref{fig:image_vs_pixel} or the 3D Gaussian splats in Figure~\ref{fig:lego} need to be regularly tested and resampled if they become degenerate.  
We implement this as compute shader launched over the target parameters after each gradient descent. 
We do not explore thorougly these complementary non-gradient-based optimizations since they are orthogonal to the gradient estimation, which is our core contribution.

\section{Results}
\label{sec:results}

\subsection{Validation of the Per-Pixel Formulation}

In Section~\ref{sec:per_pixel_formulation}, we argue that using the stochastic gradient estimator of Equation~(\ref{eq:stochastic_finite_difference}) as is, with a full-image error $f()$, would not converge in high dimensions. 
This motivates our per-pixel formulation of Equation~(\ref{eq:stochastic_finite_difference_pixel}) that we expect to alleviate the dimensionality problem. 
We test this hypothesis in Figure~\ref{fig:image_vs_pixel} where we compare the \textit{full-image} approach of Equation~(\ref{eq:stochastic_finite_difference}) and the \textit{per-pixel} approach of Equation~(\ref{eq:stochastic_finite_difference_pixel}).
In this experiment, each triangle is represented by 12 parameters (3 vertices + 1 RGB color). 
The three comparisons use respectively 12288 (1K triangles), 122880 (10K triangles), and 1228800 (100K triangles) parameters.
Note that the \textit{full-image} approach is conceptually similar to the one of Fischer and Ritschel.~\shortcite{fischer2023plateau}, that also estimates the gradient via the impact of perturbations over a full-image error, the only difference being the distribution of perturbation.
As expected, optimizing with the \textit{full-image} error is slower and impractical with large numbers of parameters. 
In contrast, our \textit{per-pixel} variant scales well up to 1M+ parameters.

\subsection{Qualitative Comparison to nvDiffRast}

Our comparison baseline is \textsc{nvDiffRast}~\cite{laine2020diffrast}, the state-of-the-art differentiable rasterizer.
Note, however, that our objective is not to compete with it in terms of performance or quality.
The promise of our method is to provide a simple-to-implement, cross-platform, and dependency-free alternative that can be incorporated into an existing rasterization engine.
Still, it is interesting to investigate how both methods compare. 
To do that, we reproduced two \textsc{nvDiffRast} samples provided by Hasselgren et al.~\shortcite{hasselgren2021appearance} with our implementation.
We show these experiments in Figures~\ref{fig:cow} and \ref{fig:skull} and we provide a performance comparison in Table~\ref{tab:perf}.
Note that brute force performance is not a relevant measure because both methods behave differently.
Indeed, \textsc{nvDiffRast} is \textbf{slower} because of its Pytorch environment but it provides \textbf{clean gradients} that allow for an efficient gradient descent.
In contrast, our method executes \textbf{faster} within a rasterization engine but provides \textbf{noisy gradients}, which make the gradient descent less efficient.
We found out that both effects counterbalance each other and that both approaches tend to produce qualitatively similar results with the same amount of optimization time. 
These experiments hence confirm that our method can be considered as an alternative to \textsc{nvDiffRast} for these applications without suffering critical performance or quality penalty. 

\begin{table}[!h]
\begin{tabular}{@{} | c | c c | c c | @{}}
\cline{1-5} 
~ & \multicolumn{2}{c|}{Figures~\ref{fig:cow}} & \multicolumn{2}{c|}{Figures~\ref{fig:skull}} \\
\cline{1-5} 
image resolution ($W \times H$)& \multicolumn{2}{c|}{$1024^2$} & \multicolumn{2}{c|}{$1024^2$} \\
number of vertices & \multicolumn{2}{c|}{$1748$} & \multicolumn{2}{c|}{$1748$} \\
texture resolution & \multicolumn{2}{c|}{$1024^2$} & \multicolumn{2}{c|}{$512^2$} \\
total number of parameters ($d$)& \multicolumn{2}{c|}{$3150972$} & \multicolumn{2}{c|}{$791676$} \\
\cline{1-5} 
~ & \textbf{\textsc{nvDiffRast}} & \textbf{ours} & \textbf{\textsc{nvDiffRast}} & \textbf{ours} \\
image/step ($N$) & 8 & 16 & 8 & 32 \\
time/step & 33ms & 21ms & 33ms & 18ms \\
\cline{1-5} 
\end{tabular}
\caption{\label{tab:perf}Performance comparison on an NVIDIA 4090 GPU.}
\vspace{-8mm}
\end{table}

\subsection{Supported Applications}

\paragraph*{Triangles, textures and volumes}

Figures~\ref{fig:image_vs_pixel}, \ref{fig:cow} and \ref{fig:skull} showcase optimizing triangle soups, meshes, textures and volumes. 
They are straightforward applications of the implementation described in Section~\ref{sec:diffast}.

\paragraph*{Subdivision surfaces}

In Figure~\ref{fig:dancer}, we apply our method to a Catmull-Clark subdivision surface~\cite{CatmullClark78} tessellated on the fly.
We optimize the coarse control mesh and the displacement and normal maps that control the final appearance. 
To support this application, we need our compute shader $\textbf{G}$ to associate each tessellated triangle to its original triangle and loop over its neighbors in the control mesh.
The subdivision data structure that we use provides a way to do this efficiently~\cite{dupuy2021}.

\paragraph*{Physically based shading}

Figure~\ref{fig:teaser} showcases a subdivision surface (same algorithm as the one of Figure~\ref{fig:dancer}) with physically based shading using roughness, metallicity, albedo, height and normal maps.

\paragraph*{3D Gaussian splats}

Figure~\ref{fig:lego} shows an optimization of 3D Gaussian Splats~\cite{kerbl2023}.
Estimating the gradient is a straightforward application of our transparent geometry support explained in Section~\ref{sec:primitives} since the splats are rasterized transparent billboard with a vertex shader and a fragment shader for the shape, color and transparency.
To improve the results, we implement an additional resampling and a splat subdivision compute shader executed after each gradient descent, following Kerb et al.~\shortcite{kerbl2023}.

\clearpage
\input{SEC_6_FIGURE_ANANAS.tex}
\clearpage
\input{SEC_6_FIGURE_COW.tex}
\clearpage
\input{SEC_6_FIGURE_SKULL.tex}
\clearpage
\input{SEC_6_FIGURE_DANCER.tex}
\input{SEC_6_FIGURE_LEGO.tex}
\input{SEC_6_FIGURE_BUNNY.tex}
\clearpage

\subsection{Performance Breakdown}

In Table~\ref{table:perf_breakdown}, we provide more fine-grained performance measures showing the timings for each stage of our method for a single optimization step.

\begin{table}[!h]
\begin{tabular}{| l | c | c | c |}
 \cline{1-4} 
 & Fig.~\ref{fig:image_vs_pixel} 1K triangles & Fig.~\ref{fig:image_vs_pixel} 10K triangles & Fig.~\ref{fig:image_vs_pixel} 100K triangles \\
 \cline{1-4} 
\textbf{P} {\scriptsize ($\times128$)} & 
\phantom{aaaa} 12$\mu$s {\scriptsize ($\times128$)} & 
\phantom{aaaa} 14$\mu$s {\scriptsize ($\times128$)} & 
\phantom{aaaa} \phantom{1}54$\mu$s {\scriptsize ($\times128$)} \\
 \cline{1-4} 
$\mathcal{R}$ {\scriptsize ($\times128$)} & 
\phantom{aaaa} 28$\mu$s {\scriptsize ($\times128$)} & 
\phantom{aaaa} 38$\mu$s {\scriptsize ($\times128$)} & 
\phantom{aaaa} 126$\mu$s {\scriptsize ($\times128$)} \\
 \cline{1-4} 
\textbf{G} {\scriptsize ($\times128$)} & 
\phantom{aaaa} 13$\mu$s {\scriptsize ($\times128$)} & 
\phantom{aaaa} 14$\mu$s {\scriptsize ($\times128$)} & 
\phantom{aaaa} \phantom{1}35$\mu$s {\scriptsize ($\times128$)} \\
 \cline{1-4} 
\textbf{D} & 
\phantom{aaaa} 21$\mu$s \phantom{{\scriptsize ($\times128$)}} & 
\phantom{aaaa} 21$\mu$s \phantom{{\scriptsize ($\times128$)}} & 
\phantom{aaaa} 105$\mu$s \phantom{{\scriptsize ($\times128$)}} \\
\cline{1-4} \cline{1-4}
Sum & 
\phantom{aaaa} 6.8ms \phantom{{\scriptsize ($\times128$)}} & 
\phantom{aaaa} 8.4ms \phantom{{\scriptsize ($\times128$)}} & 
\phantom{aaaa} \phantom{1}27ms \phantom{{\scriptsize ($\times128$)}}\\
\cline{1-4}
\end{tabular}
\vspace{2mm}
\caption{\label{table:perf_breakdown} Performance breakdown for the results of Figure~\ref{fig:image_vs_pixel}.
In this experiment, we accumulate $N=128$ stochastic gradient estimates before computing a gradient descent step (noted \textbf{D} in the table).
}
\end{table}

\section{Conclusion}

We have proposed a method to transform a non-differentiable rasterizer into a differentiable one.
Our experiments have shown that our transformed rasterizer supports the same applications as state-of-the-art differentiable rasterizers without critical performance or qualitative penalty.
We successfully used it to optimize triangles, meshes, subdivision surfaces, textures, physically based materials, volumes, and 3D Gaussian splats. 

However, we do not position our method as a replacement for other state-of-the-art differentiable rasterizers. 
Our objective is to bring the benefits of differentiable rasterization to an audience that already possesses a (non-differentiable) rasterization engine and has workflow or platform constraints that prevent using existing differentiable rasterizers. 
Our method makes it possible to enjoy the possibilities of differentiable rasterization for 3D assets optimization, within the existing engine. 
We believe that game developers who wish to optimize gaming assets withing their existing workflow will be interested in our method.

\bibliographystyle{ACM-Reference-Format}
\bibliography{bibliography}

\appendix

\end{document}